\documentclass{article}
\usepackage[utf8]{inputenc}
\usepackage{amsfonts, amsmath, amsthm, amssymb} % For math fonts, symbols and environments
\usepackage{graphicx} % Required for including images
\usepackage[super]{nth}
\usepackage{hyperref} % For hyperlinks in the PDF
\usepackage{multirow,tabu} % for table
\usepackage{array}%hide table columns
\newcolumntype{H}{>{\setbox0=\hbox\bgroup}c<{\egroup}@{}}%hide table columns
\usepackage{chngcntr}
\usepackage{bbm}
\usepackage{authblk}
\usepackage{color}
\usepackage{float} 
\usepackage[backend=biber,style=authoryear, sorting=nyt,sortcites=true,giveninits=true,maxbibnames=9,natbib ,uniquename=false,maxcitenames=2,uniquelist=false]{biblatex}
\renewbibmacro*{volume+number+eid}{%
  \printfield{volume}%
  \setunit*{\addnbspace}% NEW (optional); there's also \addnbthinspace
  \printfield{number}%
  \setunit{\addcomma\space}%
  \printfield{eid}}
\DeclareFieldFormat[article]{number}{\mkbibparens{#1}}
\title{{Statistical learning for train delays and influence of winter climate and atmospheric icing}}
\author[1]{Jianfeng Wang\thanks{Corresponding author : Jianfeng Wang\\
\hspace*{4.3 mm} Email: jianfeng.wang@umu.se}}

\author[2]{Roberto Mantas Nakhai}
\author[1]{Jun Yu}
\affil[1]{Department of Mathematics and Mathematical Statistics, Umeå University, Sweden}
\affil[2]{Department of Computer Science, Electrical and Space Engineering, Luleå University of Technology, Sweden}
\date{}

\begin{document}

\maketitle
\begin{abstract}
{This study investigated the climate effect under consecutive winters on the arrival delay of high-speed passenger trains in northern Sweden. Novel statistical learning approaches, including inhomogeneous Markov chain model and stratified Cox model, were adopted to account for the time-varying risks of train delays. The inhomogeneous Markov chain modelling for the arrival delays has used several covariates, including weather variables, train operational direction, and findings from the primary delay analysis through stratified Cox model. The results showed that the weather variables, such as temperature, snow depth, ice/snow precipitation, and train operational direction, significantly impact the arrival delay. The performance of the fitted inhomogeneous Markov chain model was evaluated by the walk-forward validation method. The averaged mean absolute errors between the expected rates and the observed rates of the arrival delay over the train line was obtained at the level of 0.088, which
implies that approximately 9\% of trains may be misclassified as having arrival delays by the fitted model at a measuring point on the train line.}
\end{abstract}

\section{Introduction}
{The arrival delay of passenger trains is one of the operational indicators that travellers mostly care about.} It measures the delay in terms of arrival time at a measuring spot. {It affects the degree of traveller's dependence and choice among possible transportation modes.} To railway operation companies, it is a key criterion in order to minimise the costs and increase the reliability of the railway operation.  {However, many factors can cause an arrival delay}. Among others, the weather is one major factor that can affect the punctual rate of trains, especially the high-speed passenger train, since this type of train has a higher priority on the train line, which reduces non-climate effects to a great extent. And it often travels longer distances, which makes the weather effect on the train more prominent. This study focuses on investigating how arrival delays of the high-speed passenger trains are affected by the winter climate. {The study region lies in northern Sweden, representing a typical region with a harsh winter climate, such as coldness, heavy snow, and ice/snow precipitation. } Such climate can cause arrival delays to railway transportation, which leads to inevitable impacts on society operations. {Besides the arrival delay, another commonly used measurement in train operation has a close relationship with the arrival delay, i.e. the primary delay.} It measures the increment in delay within two consecutive measuring spots in terms of running time. The primary delays in previous running sections on a train line can lead to arrival delays in the following running sections.

There is no unique criterion to define the time limits of arrival delays and primary delays in the world \citep{Yuan2007}. According to Swedish Transport Administration (STA), arrival delay occurs if a train arrives at one measuring spot five minutes later than its schedule  and a delay of three minutes or more in terms of running time is considered a primary delay. The STA's criterion is used in the study.

{Several studies concerning train performance have been conducted.} \citet{Yuan2007} used probability models based on blocking time theory to estimate the knock-on delays of trains caused by route conflicts and late transfer connections in stations. In \citet{MURALI2010483}, the authors modeled travel time delay as a function of the train mix and the network topology. \citet{LESSAN20191214} proposed a hybrid Bayesian network model to predict arrival and departure delays in China. \citet{Huang8883805} pointed out in their paper that arrival delay was highly correlated to capacity utilisation of the train line. In a more recent study, \citet{HUANG2020338} applied a Bayesian network to predict disruptions and disturbances during train operations in China. In addition to those, a few studies investigated weather impacts on train performance. \citet{Xia2013} fitted a linear model and showed that weather variables like snow, temperature, precipitation, and wind had significant effects on the punctuality of trains in the Netherlands. \citet{BRAZIL201769} used a simple multiple linear regression model and demonstrated that weather variables, such as wind speed and rainfall, can have a significantly negative impact on arrival delays in the Dublin area's rapid transit rail system. A machine learning approach was used to create a predictive model to predict the arrival delay at each station for a train line in China with the help of weather observations in \citet{Wang2019}. \citet{Ottosson} used negative binomial regression and a zero-inflated model and showed that weather variables, such as snow depth, temperature and wind direction, had significant effects on the train performance. Two recent studies by \citet{wang2020effects, wang2021train} applied non-stratified/stratified Cox model and homogeneous/inhomogeneous Markov chain model to analyse winter effects on the primary delay and arrival delay, respectively. Thereinto, the inhomogeneous models have been proved to be better than the homogeneous counterparts in \citet{wang2021train}. The authors treated primary delay as recurrent time-to-event data, and the transitions between states arrival delay and punctuality as a Markov chain. {Common limitations in the two studies are, however, that the authors used one single winter season which contained limited amount of data and may result in biased estimates, and no evaluation was performed to the fitted model, thus the reliability of the statistical models could be a potential issue.}

 In this study, we analyse the effect of consecutive winters on the arrival delay and the primary delay of the high-speed passenger trains in northern Sweden. {Novel statistical learning approaches including inhomogeneous Markov chain model and stratified Cox model were adopted to account for the time-varying risks of train delays and model performance assessment.} Because of the close relationship between primary delay and arrival delay, we first use stratified Cox model to analyse the primary delay. The estimated survival curve of the primary delay is instrumental in identifying the changes in transition intensity between arrival delay and punctuality. Based on the findings from the analysis of primary delay, an inhomogeneous Markov chain model is setup to analyse the winter effect on the arrival delay. Afterwards, the performance of the fitted inhomogeneous Markov chain model is evaluated using an expanding window walk-forward validation method with mean absolute error (MAE) to assess the accuracy of its prediction capability. Therefore, the main improvement of this study is that the length of analysis period is doubled and the model evaluation is conducted, which gives a more comprehensive understanding on the model performance. 
 
The paper is organised as follows. In Section 2, we introduce the statistical models in detail. Data and evaluation method are described in Section 3. Section 4 is reserved for results. Section 5 is devoted to the conclusion and discussion.

\section{Statistical modelling}
In the section, stratified Cox model and inhomogeneous Markov chain model are introduced for the analysis of the primary delay and arrival delay, respectively. 
\subsection{Stratified Cox model}
 \citet{PRENTICE} proposed a stratified Cox model. It is an extension of Cox
models in \citet{Cox,Andersen1982}. The stratified Cox model is commonly used for modelling recurrent events in survival analysis. It is used in this study to analyse how the time-dependent weather variables affect the recurrent primary delay of the train. The model assumes that the hazard function of a primary delay is correlated to its preceding primary delays through an event-specific baseline hazard function. The stratified Cox model is given by

\begin{equation}
\label{coxmodel}
   h_{ij}(t)  = h_{0j}(t)\exp{(\boldsymbol{\beta}^T \mathbf{x}_{ij}(t))},
\end{equation}
where $h_{ij}(t)$ represents the hazard function for the $j$th primary delay of
the $i$th train at time $t$, $h_{0j}(t)$ is an event-specific baseline hazard and $j$ is a stratification variable, e.g. $h_{01}(t)$ is a common baseline hazard of the first primary delay for each train, $\boldsymbol{\beta}$ is an unknown  coefficient vector to be estimated and $\mathbf{x}_{ij}(t)$ represents a covariate vector for the $i$th train and the $j$th primary delay at time $t$.

The coefficients are estimated by maximising the partial likelihood. It takes the conditional probabilities of the occurrence of primary delays over all the trains into account. The likelihood is given by
\begin{equation}
    \label{partialLikli}
    L(\boldsymbol{\beta})  = \prod_{i = 1}^n \prod_{j = 1}^{k_i} \left(\frac{\exp{( \boldsymbol{\beta}^T \mathbf{x}_{i}(t_{ij})))}}{\sum_{l \in R(t_{ij})} \exp{( \boldsymbol{\beta}^T \mathbf{x}_{l}(t_{ij})})}\right)^{\delta_{ij}},
\end{equation}
where $j$ is the event index with $k_i$ being the train-specific maximum number of events, $\mathbf{x}_{i}(t_{ij})$ denotes the covariate vector for the $i$th train at the $j$th event time $t_{ij},$ $\delta_{ij}$ is an event indicator variable taking $1$ for the $j$th event of the $i$th train and $0$ for censoring, $R(t_{ij}) = \{l, l=1,\cdots,n: t_{l(j-1)}< t_{ij}\leq t_{lj}\}$ is a group of trains that are at risk for the $j$th event at time $t_{ij}$. 

With the fitted model, the hazard function, $\hat{h}_{ij}(t)$, for the $j$th primary delay of train $i$ can be estimated. The corresponding survival function, $\hat{S}_{ij}(t),$ can also be calculated using the estimated hazard function through $\hat{S}_{ij}(t)=\exp\left(-\int_0^t \hat{h}_{ij}(x)\, \mathrm{d}x\right)$.  $\hat{S}_{ij}(t)$ gives the probability that train $i$ has not suffered the $j$th primary delay up to time $t$. 

\subsection{Inhomogeneous Markov chain model}
Let $\{Y(t), t\ge 0\}$ denote a continuous-time Markov chain, and $Y(t)$ takes values over a countable state space. $P(Y(t)=s)$ is the probability of chain $Y$ at state $s$ at time $t$. The transition probability of the chain moving from state $r$ at time $t$ to state $s$ at time $t+u$ is denoted by conditional probability $p_{rs}(t, t+u)=P(Y(t+u) = s | Y(t) = r) $. The instantaneous change from state $r$ to state $s$ at time $t$ is controlled by transition intensity, $q_{rs}(t)$, through transition probability
\begin{equation}
\label{inte}
    q_{rs}(t) = \lim_{\Delta t \to 0} P( Y(t+\Delta t) = s |Y(t)=r)/\Delta t.
\end{equation}
With these definitions, a Markov chain can be used to describe transitions between train states (arrival delay/punctuality) on a train line, where the time $t$ refers to the running distance of a train under the context instead of clock time, since the running distance is more meaningful in practice. The $q_{rs}(t)$ of a $k$ states process forms a $k\times k$ transition intensity matrix $Q(t)$, whose rows sum to zero, so that the diagonal entries are defined by $q_{rr}(t) = -\sum_{s\neq r}q_{rs}(t)$.

A homogeneous Markov chain in time means that the transition intensities in $Q(t)$ are constant over time, and the transition probability from one state to another depends only on the time difference between the two-time points, i.e. 

\begin{equation}
    \label{mark2}
    P( Y(t+u) = s |Y(t) = r) = P(Y(u) = s | Y(0) = r).
\end{equation}

In analogous to the transition intensity matrix $Q(t)$, the entry of a transition probability matrix $P(t,t+u)$ is the transition probability $p_{rs}(t,t+u)$. For a homogeneous process, the relationship between transition intensity matrix $Q$ and transition probability matrix $P(t, t+u)$ is specified by the Kolmogorov differential equations \citep{Cox1977}. More specifically, the transition probability matrix can be calculated by taking the matrix exponential of the transition intensity
matrix 
\begin{equation}
\label{pro}
    P(t,t+u)=P(u) = \text{Exp}(uQ).
\end{equation}

% In a homogeneous Markov chain model, to take account of the effect of covariates, a Cox like model was proposed by \citet{Marshall1995}
% \begin{equation}
% \label{explo}
%     q_{rs} = q_{rs}^{(0)} \exp{(\boldsymbol{\beta}_{rs}^T \mathbf{x}_{rs})},
% \end{equation}
% where $q_{rs}^{(0)}$ is a baseline transition intensity from state $r$ to state $s$ when all covariates are zero and $\mathbf{x}_{rs}$ is a covariate vector under the corresponding transition.  The value $\exp{(\beta_{rs})}$, where $\beta_{rs}$ is one element of the vector $\boldsymbol{\beta}_{rs},$ reflects how the corresponding covariate affects the hazard ratio given that all other covariates are held constant. More specifically, $\exp{(\beta_{rs})}>1$ indicates the transition intensity from $r$ to $s$ increases as the value of the covariate increases, $\exp{(\beta_{rs})}<1$ indicates the transition intensity decreases as the value of the covariate increases, while $\exp{(\beta_{rs})}=1$ implies the covariate has no effect on the transition intensity.

Since the weather variables along the train line vary over time and are only available at the measuring spots at the same times as the states of the Markov chain are derived, the approximate effects of the weather variables between two consecutive spots are estimated using the average of the two weather data points at the spots. In addition, the estimated survival curve of the primary delay also helps to identify the changing points of the transition intensity. These two factors lead to an inhomogeneous Markov model with the piece-wise constant transition intensity, i.e.
\begin{equation}
\label{exten}
    q_{rs}(t) = q_{rs}^{(0)} \exp{(\boldsymbol{\beta}_{rs}^T \mathbf{x}_{rs}(t)+z_{rs}\mathbbm{1}_{\{t\ge t_0\}})},
\end{equation}
where $q_{rs}^{(0)}$ represents the baseline transition intensity from state $r$ to $s$, $\boldsymbol{\beta}_{rs}$ is a coefficient vector to be estimated, $\mathbf{x}_{rs}(t)$ is a covariate vector from state $r$ to $s$ at time $t$, $z_{rs}$ is another coefficient to be estimated and $\mathbbm{1}$ is an indicator variable taking value 1 if $t\ge t_0$, otherwise, 0. The indicator variable reflects where the transition intensity changes, and more than one indicator function can be added to the model if there are more than one changing points on the train line that are detected in the estimated survival curve.  
% Similar to $\exp{(\beta_{rs})}$, the value $\exp{(z_{rs})}$ is the hazard ratio of intensities between $t\ge t_0$ and $t<t_0$ for the transition from $r$ to $s$. 

The coefficient vectors $\boldsymbol{\beta}_{rs}$ and $z_{rs}$ as well as the transition intensity matrix $Q(t)$ can be estimated by maximising the likelihood. For instance, the likelihood for (\ref{exten}) is 

\begin{equation}
    \label{logts}
    L(Q(t)) =  \sum_{Y(t_{i,t_0})\in D}\prod_{i = 1}^n \prod_{j = 1}^{c_i} p_{Y(t_{i,j}),Y(t_{i,j+1})}(t_{i,j + 1}- t_{i,j}),
\end{equation}
where $j$ represents a time index either at a measuring spot or at a changing point of the transition intensity with $c_i$ being the total number of the indices for train $i$ on the train line, $Y(t_{i,j})$ represents the $j$th state of the $i$th train at time $t_{i,j}$ and the transition probability is evaluated at the time difference $t_{i,j + 1}- t_{i,j}$. In the case that $t_0$ is a point between two measuring spots where no state can be derived, $Y(t_{i,t_0})$ is treated as censored data and a set $D$ contains all possible states of $Y(t_{i,t_0})$, which results in the summation in the likelihood.

With the fitted inhomogeneous Markov chain model, transition probability matrix $\hat{P}(t)$ can be estimated for any operational interval of interest on the train line for prediction purpose.

\section{Data and method}
This section provides a description of the train data and covariates used in the study, as well as the evaluation method of the prediction capability of the fitted inhomogeneous Markov chain model. 

\subsection{Train data}
 High-speed passenger trains between Stockholm and Umeå in the northern region of Sweden are investigated. It is a type of train with a top speed of between 200 to 250 km/h. The total length of the train line is 711 km and the planned drive time is 6.5 hours. On the train line it comprises 116 measuring spots. At each spot, a train's departure and arrival times are recorded. The lengths of any two consecutive measuring spots vary from 0.3 km to 15 km. Table \ref{table:1} lists the relevant necessary variables in the dataset.
\begin{table}[H]
\centering
\caption{Key variables}
\medskip
\begin{tabular}{  m{11em} | m{11cm}  } \hline\hline
\textbf{Variables}  & \textbf{Description} \\ \hline
Train Number & Identification number of each train in the trip  \\ \hline
Initial station & Umeå or Stockholm  \\ \hline
Arrival spot & Name of arrival measuring spot  \\ \hline
Departure spot & Name of departure measuring spot \\ \hline
Departure date &  The departure date for a train at a location  \\ \hline
Arrival date &  The arrival date for a train at a location \\ \hline
Section Length  & Length between two consecutive measuring spots (km) \\ \hline
Planned departure time  & The planned departure time at a measuring spot (hh:mm)  \\ \hline
Planned arrival time  & The planned arrival time at a measuring spot (hh:mm)  \\ \hline
Actual departure time  & The Actual departure time at a measuring spot (hh:mm) \\ \hline
Actual arrival time  & The Actual arrival time at a measuring spot (hh:mm)  \\ \hline\hline
\end{tabular}
\label{table:1}
\end{table}

\begin{figure}[H]
    \centering
    \includegraphics[scale=0.37]{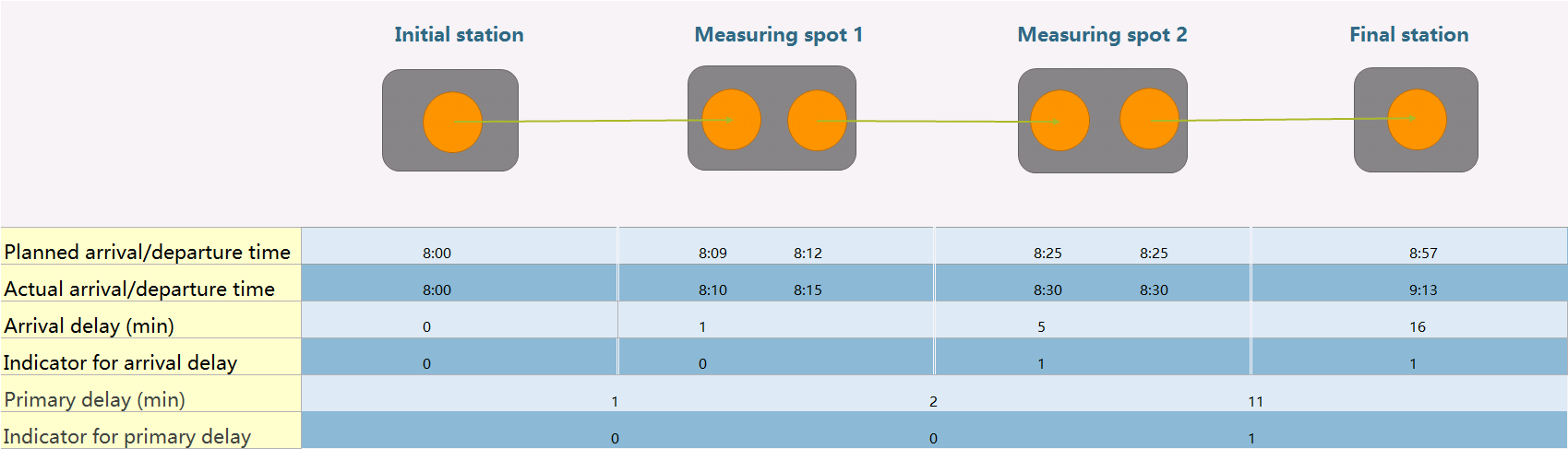}
    \caption{Illustration of a train run with derived indicators for primary and arrival delays}
    \label{pic:timetable}
\end{figure}

To fit the two statistical models to the train data, the dataset should include the following variables, i.e. each data record contains one departure spot, its subsequent arrival spot, length of each measuring spot from the initial station, and indicator variables of primary delay and arrival delay, 0/1, as well as corresponding covariates. To derive the indicator variables, one needs to calculate the running time difference (Actual arrival time$-$Actual departure time)$-$(Planned arrival time$-$Planned departure time), and arrival time difference (Actual arrival time$-$Planned arrival time), compared to the schedule. With the calculated numbers, the values for the two indicator variables can be obtained, i.e. 1 stands for a primary/arrival delay, 0 otherwise. Figure \ref{pic:timetable} is an example to illustrate how to derive the indicator variables on a train line with four measuring spots. Besides, to take account of the operational direction of a train and to be able to detect the difference caused by it in the analysis, a binary direction variable is created as a covariate in the models based on the variable Initial station, i.e. 1 represents a train running from Umeå to Stockholm and 0 denotes from Stockholm to Umeå.

\subsection{Weather data} 
\begin{figure}[H]
    \centering
    \includegraphics[scale=0.7]{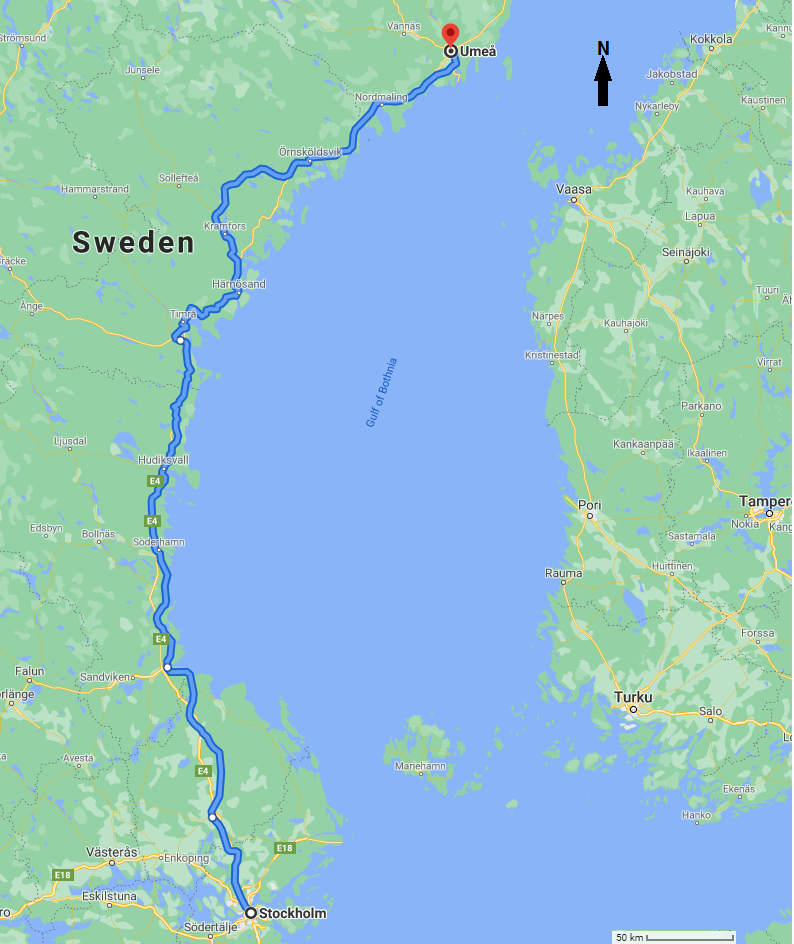}
    \caption{Train line in the region with simulated WRF data}
    \label{pic:weather}
\end{figure}

 In Sweden, December -- February is a typical winter season, thus the consecutive winters, i.e.  December 2016 -- February 2017 and December 2017 -- February 2018, are chosen as the analysis period. {As for the weather variables, we use the simulated data from the Weather Research and Forecasting (WRF) model,  instead of the actual data from the meteorological observations, since the latter has feeble spatial resolutions, where the distances between the nearest meteorological station and measuring spot range from 17 km to 24 km \citep{Ottosson}. Moreover, there is no observational data for atmospheric icing from the meteorological stations. The WRF model is a numerical weather prediction system with satisfactory accuracy that is commonly used for operational and research purposes. }The model's reliability has been presented in several studies \citep{wangbayesian, wangdownscal,Mohan,Cassano}. The WRF model simulates weather variables over grids at each single time point. Various spatial resolutions and temporal resolutions are available in the simulation setting. This study uses the spatial resolution of $3\times3$ km and the temporal resolution of 1 hour. Figure \ref{pic:weather} draws the simulation region and the train line under investigation.

Table \ref{table:2} lists the weather variables under investigation. These variables are chosen because they are shown to have impacts on the train operation in a single winter \citep{wang2021train,wang2020effects,Ottosson}.

\begin{table}[H]
\centering
\caption{The weather variables under investigation}
\medskip
\begin{tabular}{ m{10em} | m{10cm}  } 
\hline\hline
\textbf{Variables}  & \textbf{Description} \\ 
\hline
Temperature  & The temperature at 2 meters above the ground ($^\circ$C) \\  \hline 
Humidity  & Relative Humidity at 2-meters (\%)   \\  \hline
Snow depth & The snow depth (cm)  \\  \hline
Ice/snow precipitation  &  Hourly accumulated ice/snow (mm) \\  \hline\hline

\hline
\end{tabular}

\label{table:2}
\end{table}
 The measuring time of the trains needs to be rounded to the closest hour in order to match every measuring spot on the train line with the nearest grid point by time. The averages of the weather variables within any two consecutive spots are calculated and used in the two models. Since many ice/snow precipitation values are zero on the train line, a categorical variable is used instead of the continuous variable, i.e. 1 if ice/snow precipitation is not zero, 0 otherwise.

\begin{figure}[H]
    \centering
    \includegraphics[width=18cm,keepaspectratio]{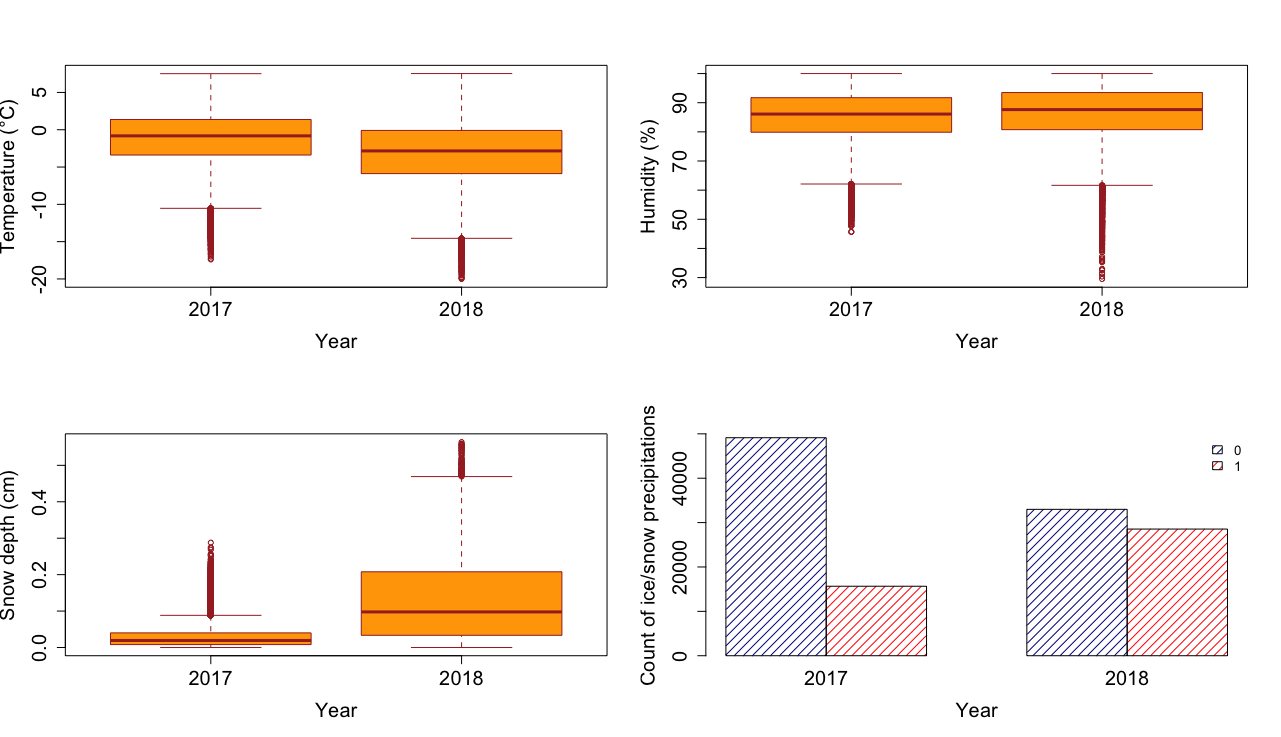}
    \caption{Illustration of the four weather variables in each winter}
    \label{pic:des}
\end{figure}
To have an intuitive understanding of the four weather variables and to visualise the differences between the two winters, three-box plots and one bar chart are illustrated in Figure \ref{pic:des}. The differences between the two winters can be clearly observed. In general, winter 2018 has a lower temperature, the extremely lower humidity, deeper snow and more ice/snow precipitations compared with winter 2017.

\subsection{Evaluation method}
To evaluate the performance of the fitted inhomogeneous Markov chain model constructed upon the covariates and information extracted from the stratified Cox model, the prediction ability of the fitted model is assessed. For this purpose, an expanding window walk-forward validation method is applied \citep{s21072430,KOHZADI1996169}. It is a sliding window-based method with the training data expanding gradually in ascending order of time. With this method, one actually does multiple-time model training and validations, then the averaged performance of the model can be obtained, which provides a more robust measure of the model performance than the one-time model training and validation.

\begin{figure}[H]
    \centering
    \includegraphics[width=14cm,keepaspectratio]{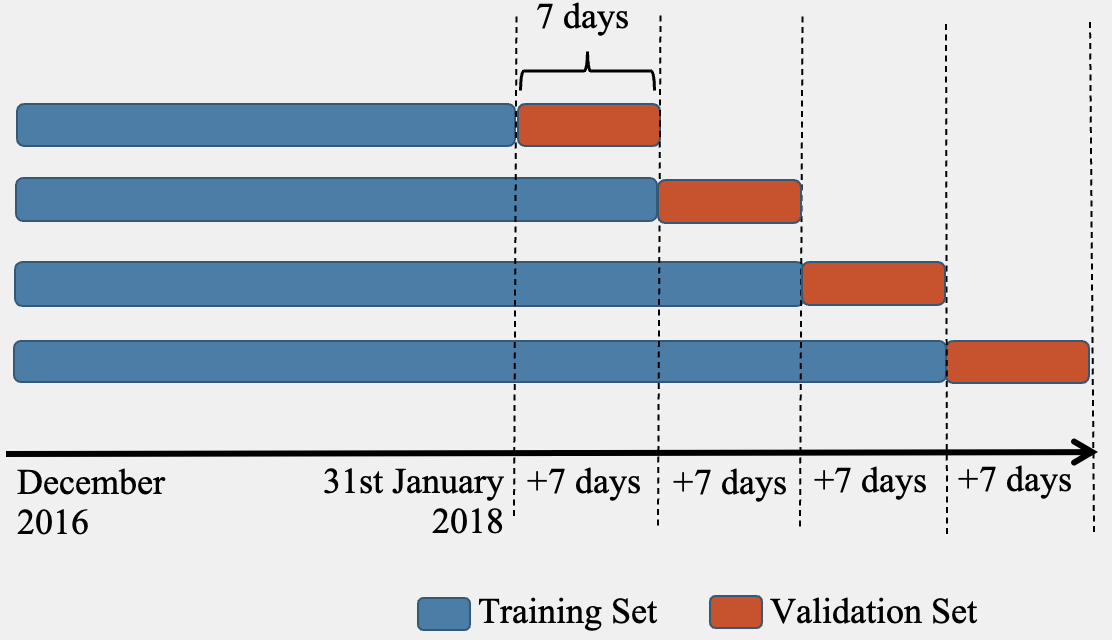}
    \caption{Illustration of expanding window walk-forward validation method.}
    \label{pic:walk}
\end{figure}

A four-time model trainings and validations scheme of the study is illustrated in Figure \ref{pic:walk}. Each time the validation period contains 7 days and the training period of the first time is from December 2016 to  \nth{31} January 2018 and the corresponding validation period is the first 7 days in February 2018. Afterwards, the new training period in the next step sequentially expands the previous period with 7 days and the new validation period moves forward to contain the next 7 days until the last validation period which includes the last 7 days of February.

{Evaluation of model performance is conducted by comparing the expected rate of arrival delays predicted from the fitted model and the observed rate of arrival delays from the validation data.} For this purpose, MAE is selected for the comparison at a number of pre-specified evaluation points, $\{50,100,150,\dots, 700\}$, on the train line over each validation period. {The observed rate at a certain evaluation point is the ratio of the number of trains having arrival delays to the total number of trains at the moment in the validation data.} The expected rate at one evaluation point is the predicted rate of arrival delay from the fitted model at the point. MAE is an useful quantity commonly used to measure forecast error in time series analysis \citep{HYNDMAN2006679, CALI2018240}. The formula of MAE for the $k$th model training and validation step is expressed as 
\begin{equation}
    \label{mae}
    \text{MAE}_k = \frac{1}{n}\sum_{i=1}^{n}\left | \hat{r}_{ki} - r_{ki} \right |,
\end{equation}
where $n=14$ is the total number of the evaluation points, $\hat{r}_{ki}$ denotes the expected rate of arrival delay at the evaluation point $i$ which is derived from the predicted transition probability matrix from the fitted model, and $r_{ki}$ represents the observed rate of arrival delay at $i$ for the $k$th model training and validation. The averaged MAE over the four-time model trainings and validations, $\text{MAE}_k$ with $k=1,2,3,4,$ can be calculated as an overall quantity for the measurement of the model performance.        

% \subsection{Analysis tool}
% R is the software used for data processing and modelling. Specifically, the package \textit{survival} is used for the stratified Cox model and the package \textit{msm} is used for the inhomogeneous Markov chain model. 

\section{Results}
\subsection{Stratified Cox model}
Table \ref{table:solocox} illustrates the estimates from the fitted stratified Cox model with 95\% confidence intervals (CIs) and $p$-values. Snow depth and ice/snow precipitation are the two variables that have significant effects on the occurrence of the primary delay. To be specific, as snow depth increases 1 cm, the hazard increases 1.4\%, and  as ice/snow precipitation increases 1 mm, the risk rises 22.5\%.

\begin{table}[H]
\caption{Estimates from the fitted stratified Cox model }
\medskip
\centering
\begin{tabular}{  m{10em} |  m{2.5cm}| m{2cm}| m{2cm}|    m{2cm} } 
\hline\hline
\textbf{Predictor}  &\textbf{Hazard ratio} & \textbf{CI: Lower}& \textbf{CI: Upper} &  \textbf{$p$-value}\\ \hline
Direction  & 1.021&0.915&1.140& 0.7063     \\\hline 
Temperature  & 0.989&0.973&1.004& 0.1618     \\\hline 
Humidity & 1.001&0.9937&1.008&  0.8593    \\\hline 
Snow depth & 1.014&1.009&1.020 &0.0000     \\\hline 
Ice/snow precipitation & 1.225&1.088&1.379 &0.0008    \\\hline \hline

\end{tabular}

\label{table:solocox}
\end{table}

\begin{figure}[H]
    \centering
    \includegraphics[scale=0.35]{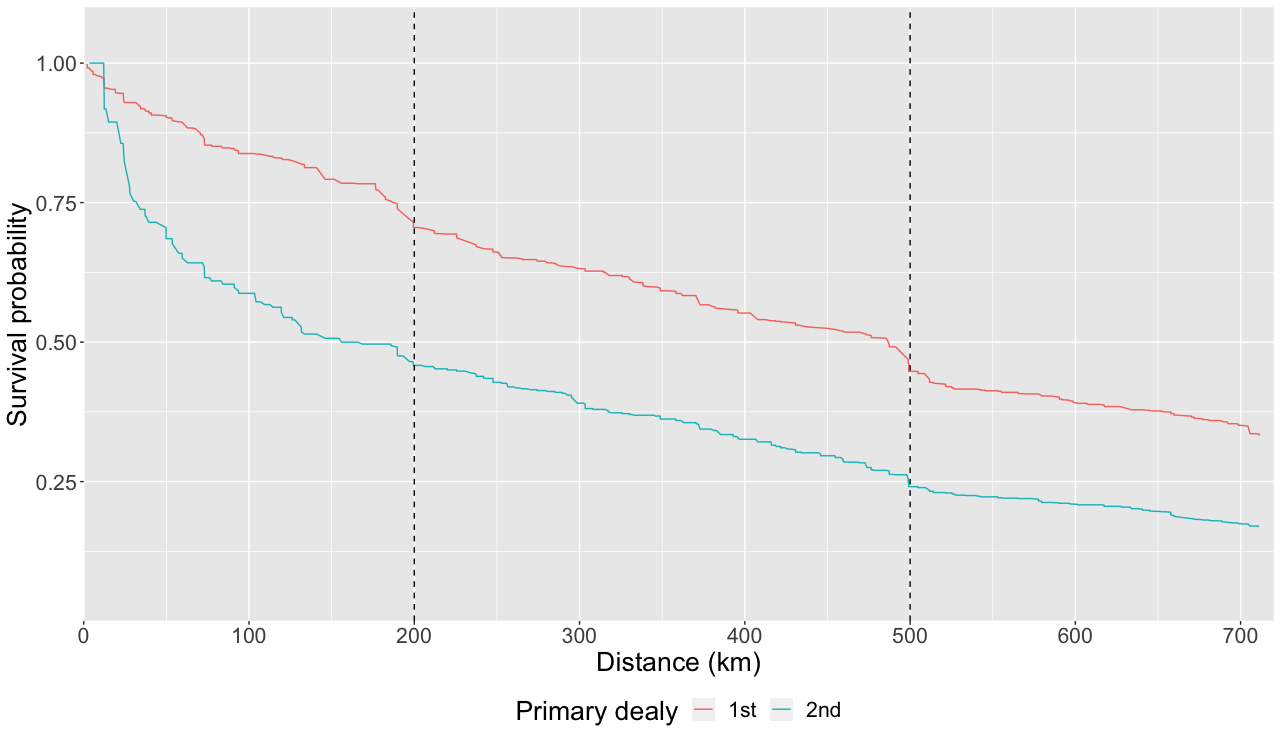}
    \caption{Survival probabilities for the first two primary delays}
    \label{pic:sur}
\end{figure}

A survival plot from the fitted model is drawn to show how survival probabilities of a train running from Umeå to Stockholm vary between the first and second occurrence of primary delays in Figure \ref{pic:sur}. The survival curves for the higher orders of primary delays are not shown due to the data deficiency. The curves are plotted under the condition with the average of temperature, humidity and snow depth among the whole data together with ice/snow precipitation, i.e. the temperature is $-1.2^\circ$C, humidity is 85\%, snow depth is 3 cm and ice/snow precipitation is 1. There are two noticeable reductions at 200 km and 500 km on the two curves, respectively, in the figure.

\subsection{Inhomogeneous Markov chain model}
As indicated in Figure \ref{pic:sur}, there exist two substantial reductions on the curves under the average weather condition. Thus it is reasonable to assume the transition intensity varies at the distance 200 km and 500 km, respectively. Therefore, two indicator variables are set in the inhomogeneous Markov chain (\ref{exten}), i.e. $\mathbbm{1}_{\{t\in [200,500) \}}$ and $\mathbbm{1}_{\{t\ge 500\}}$. Table \ref{table:markov} and \ref{table:markov2} present the hazard ratios from the inhomogeneous Markov chain model with 95\% CIs and $p$-values. The direction, temperature, snow depth and ice/snow precipitation significantly impact both transitions from punctuality to arrival delay and from arrival delay to punctuality.  Specifically, in Table \ref{table:markov}, the transition intensity from punctuality to arrival delay decreases 41.4\% for the train departures from Umeå, decreases 4.2\% as the temperature increases $1^\circ$C, increases 2.6\% as the snow depth increases 1 cm, and the occurrence of ice/snow precipitation increases the transition intensity 14.2\%. On the other side, Table \ref{table:markov2} indicates that the transition intensity from delayed to punctual states decreases 24.9\% if the train starts from Umeå, as the temperature increases $1^\circ$C, the transition intensity from delayed to punctual states increases 1.7\%, as the snow depth increases 1 cm, the transition intensity decreases 1.6\%, and the occurrence of ice/snow precipitation decreases the transition intensity  23.5\%. Thereinto, humidity is the only non-significant variable in the two tables, and the direction has smaller hazard ratios in both of the two tables, which implies the train running from Umeå is not easy to change its state compared to the one from Stockholm.

\begin{table}[H]
\centering
\caption{Hazard ratios from punctuality to arrival delay}
\medskip
\begin{tabular}{  m{10em} | m{2.5cm}|  m{2cm}| m{2cm}|m{2cm}} 
\hline\hline
\textbf{Predictor}  & \textbf{Hazard Ratio} & \textbf{CI: Lower}& \textbf{CI: Upper}& \textbf{$p$-value}   \\ \hline
Direction & 0.586&0.525&0.653&$<0.0001  $  \\\hline 
Temperature & 0.958&0.943&0.973&$<0.0001  $  \\\hline 
Humidity &  0.997 &0.991&1.003&0.3757 \\\hline 
Snow depth & 1.026& 1.020&1.031 &$<0.0001  $ \\\hline 
Ice/snow precipitation & 1.142&1.015&1.285 &0.0270  \\\hline \hline

\end{tabular}

\label{table:markov}
\end{table}

\begin{table}[H]
\centering
\caption{Hazard ratios from arrival delay to punctuality}
\medskip
\begin{tabular}{  m{10em} | m{2.5cm}|  m{2cm}| m{2cm}|m{2cm}} 
\hline\hline
\textbf{Predictor}  & \textbf{Hazard Ratio} & \textbf{CI: Lower}& \textbf{CI: Upper}&\textbf{$p$-value}    \\ \hline
Direction & 0.751&0.663&0.850&$<0.0001  $  \\\hline 
Temperature & 1.017&1.000&1.033 &0.0471 \\\hline 
Humidity &  1.003 &0.996&1.010&0.4197 \\\hline 
Snow depth & 0.984& 0.978&0.991 &$<0.0001  $ \\\hline 
Ice/snow precipitation  & 0.765&0.671&0.872&$<0.0001  $   \\\hline \hline

\end{tabular}

\label{table:markov2}
\end{table}

Table \ref{table:intens1} and \ref{table:intens2} reflect the relation of transition intensity among the three segments on the train line. In Table \ref{table:intens1}, it indicates that the transition intensity in the second segment of the trip from punctual to delayed states is 94.7\% higher than the one in the first segment, however, there is no significant difference between the transition intensity from  delayed to punctual states in the first two segments. Table \ref{table:intens2}, shows the comparison between the third segment and the first segment. It shows that the transition intensity in the third segment of the trip from punctual to delayed states is 65.8\% higher than the one in the first segment, and transition intensity from delayed to punctual states is 28.2\% lower than the first segment. In brief, the middle segment has the highest chance of a transfer from punctuality to delay, and the last segment has the lowest probability of recovering from a delayed state.
\begin{table}[H]
\centering
\caption{Estimated Hazard ratios between segments $[200, 500)$ and $[0,200)$ }
\medskip
\begin{tabular}{m{9em} | m{2.5cm}|  m{2cm}| m{2cm}| m{2cm}} 
\hline\hline
\textbf{Predictor}  & \textbf{Hazard Ratio} & \textbf{CI: Lower}& \textbf{CI: Upper}& \textbf{$p$-value}    \\ \hline
Punctuality - delay & 1.947&1.690&2.242& $<0.0001  $ \\\hline 
Delay - punctuality & 1.038&0.870&1.238&$0.6951$   \\\hline \hline
\end{tabular}

\label{table:intens1}
\end{table}

\begin{table}[H]
\centering
\caption{Estimated Hazard ratios between segments $[500, \text{end})$ and $[0,200)$ }
\medskip
\begin{tabular}{m{9em} | m{2.5cm}|  m{2cm}| m{2cm}| m{2cm}} 
\hline\hline
\textbf{Predictor}  & \textbf{Hazard Ratio} & \textbf{CI: Lower}& \textbf{CI: Upper}& \textbf{$p$-value}    \\ \hline
Punctuality - delay & 1.658&1.417&1.941& $<0.0001  $ \\\hline 
Delay - punctuality & 0.718&0.599&0.861&$0.0004$   \\\hline \hline
\end{tabular}

\label{table:intens2}
\end{table}

\begin{figure}[H]
    \centering
    \includegraphics[scale=0.4]{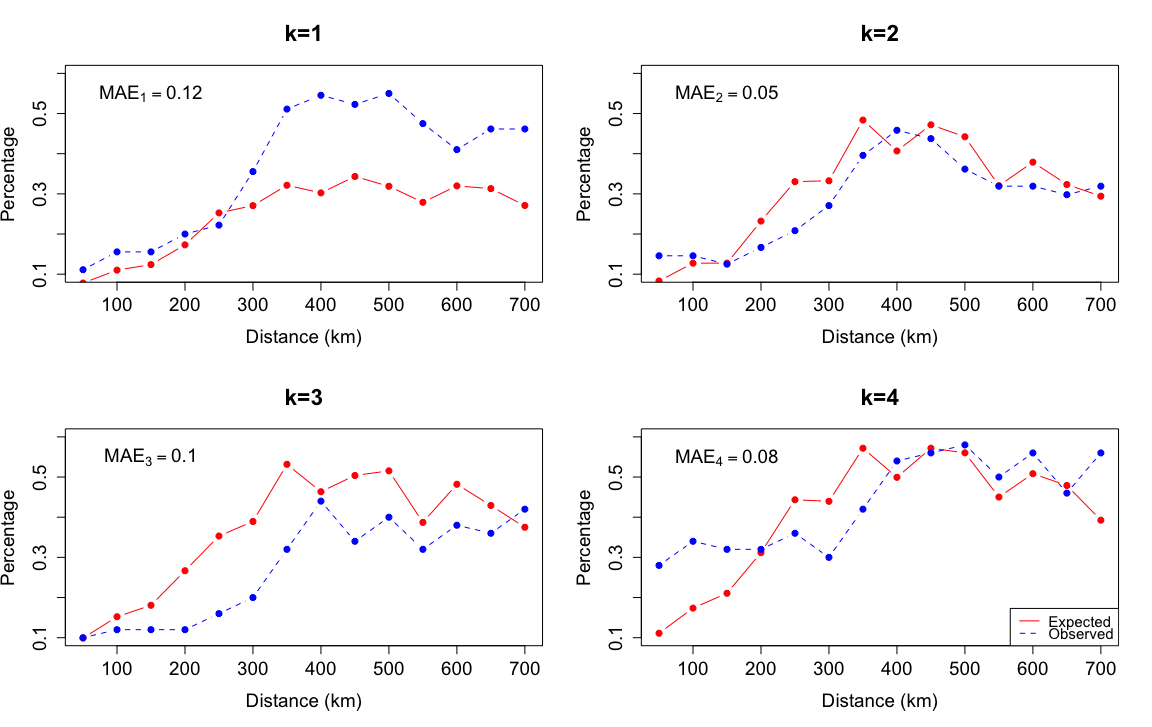}
    \caption{Comparisons between observed and expected rates of arrival delays over the trip for the four validation periods}
    \label{pic:prob}
\end{figure}

Finally, Figure \ref{pic:prob} compares observed rates of arrival delays and expected rates in each validation period according to the four-time model training and validations scheme. The largest discrepancy between the two rates occurs in the first validation period after 300 km which may be caused by specific unknown reasons for example infrastructure problems on the train line. In the second period, it has the smallest discrepancy with MAE$_2=0.05$. The average MAE over the four periods is 0.088, which implies about 9 out of 100 trains may be misclassified into a wrong state by the fitted model at a measuring point on the train line.

\section{Conclusion and discussion}

This study investigated the climate effect under consecutive winters on the arrival delay of high-speed passenger trains in northern Sweden. Novel statistical learning approaches, including inhomogeneous Markov chain model and stratified Cox model were adopted to account for the time-varying risks of train delays. The inhomogeneous Markov chain modelling for the arrival delays made use of a number of covariates including weather variables, train operational direction, and findings from the primary delay analysis through stratified Cox model. The differences of the weather variables between the consecutive winters were compared. It showed that winter 2018 was a much harsher winter than winter 2017. The estimates from the model stated that the weather variables, such as temperature, snow depth, ice/snow precipitation, and train operational direction, had significant impacts on the arrival delay. The performance of the fitted inhomogeneous Markov chain model was evaluated by the walk-forward validation method through MAE. The averaged MAE between the expected rates and the observed rates of the arrival delay over the train line was obtained at the level of 0.088. The smaller MAE indicated that the current model was able to capture the real inhomogeneous transition intensity, and it implied that approximately 9\% of trains might be misclassified as having arrival delays by the fitted model at a measuring point on the train line. In addition, since the validation periods were chosen from the harsh winter and the training periods were covered mainly by the mild winter, the performance of the fitted model was not seriously affected by such selection. Thus, it provided a shred of evidence for the robustness of the model.

To further improve the model's prediction capability, the following aspects could be taken into account in future research work: 1) more consecutive winters need to be included in the modelling part to acquire more robust statistical conclusions; 2) more train lines should be involved in the analysis to take into account the heterogeneity among train lines; 3) the inhomogeneous effect among trains should be considered by using, for example, Bayesian model or frailty Cox model \citep{vanniekerk2019new}; 4) continuously changing transition intensity with Weibull distributed time may be worth to use. It is more plausible than the piece-wise constant inhomogeneous model \citep{Titman}. Such a model can reduce the subjective factor of deciding the changing point of the transition intensity and the number of changing points in an inhomogeneous Markov chain; 5) last not least, more influential covariates and interaction effects between covariates may be incorporated in the model.

\section*{Acknowledgements}
We acknowledge EU Intereg Botnia-Atlantica Programme and Regional Council of V\"{a}sterbotten and Ostrobothnia for their support of this work through the NoICE project. We would like to thank the Swedish Transport Administration for providing the train operation data and the High Performance Computing Center North (HPC2N) and the Swedish National Infrastructure for Computing (SNIC) for providing the computing resources needed to generate the WRF data.  %We would like to thank the Editor and anonymous reviewers for their detailed and insightful comments and suggestions that helped to improve the quality of the paper.
  
\printbibliography
\end{document}